\makeatletter
\declare@file@substitution{revtex4-1.cls}{revtex4-2.cls}
\makeatother
\documentclass[twocolumn]{aastex631}
\usepackage{natbib}
\usepackage{amsmath}	% Advanced maths commands

\graphicspath{{./}{figures/}}
\begin{document}

\title{Origin of the twice ${90}^{\circ}$ rotations of the polarization angle in GRB 170114A and GRB 160821A}

\author{Xu Wang}
\author[0000-0001-5641-2598]{Mi-Xiang Lan}
\affiliation{Center for Theoretical Physics and College of Physics, Jilin University, Changchun, 130012, China; lanmixiang@jlu.edu.cn \\}

\author[0000-0001-7471-8451]{Qing-Wen Tang}
\affiliation{Department of Physics, School of Physics and Materials Science, Nanchang University, Nanchang, 330031, Jiangxi, China \\}

\author[0000-0002-6299-1263]{Xue-Feng Wu}
\affiliation{Purple Mountain Observatory, Chinese Academy of Sciences, Nanjing, 210023, China; xfwu@pmo.ac.cn \\}

\author[0000-0002-7835-8585]{Zi-Gao Dai}
\affiliation{Department of Astronomy, School of Physical Sciences, University of Science and Technology of China, Hefei, 230026, China; daizg@ustc.edu.cn \\}

\begin{abstract}

The observed twice abrupt ${90}^{\circ}$ rotations of the polarization angle (PA) in the prompt phase of gamma-ray bursts (GRBs) are difficult to be understandable within the current one-emitting-shell models. Here, we apply a model with multiple emitting shells to solve this new challenging problem. Two configurations of large-scale ordered magnetic fields in the shells are considered: toroidal and aligned. Together with the light curves and the spectral peak-energy evolutions, the twice ${90}^{\circ}$ PA rotations in GRB 170114A and GRB 160821A could be well interpreted with the multi-shell aligned magnetic fields configuration. Our numerical calculations also show that the multiple shells with the toroidal magnetic field configuration could not explain the observed twice ${90}^{\circ}$ PA rotations. An aligned magnetic field configuration in the GRB outflow usually indicate to prefer a magnetar central engine, while a toroidal field configuration is typically related to a central black hole. Therefore, the magnetar central engines for the two GRBs are favored.

\end{abstract}

\keywords{Gamma-ray bursts (629); magnetic fields (994);}

\section{Introduction}\label{intro}

Gamma-ray bursts (GRBs) are the most violent bursts in the universe, whose origin is still mysterious. There are two kinds of GRB central engines, i.e., the magnetar and the black hole. The outflow from a magnetar is a stripped wind and the magnetic field configuration (MFC) in the wind is aligned, while the jet from a black hole is powered by Blandford-Znajek mechanism \citep{Blandford_1977} and the MFC in the jet emission region is usually toroidal \citep{Spruit_2001}. Generally speaking, both the light curves and the energy spectra would depend on the strength of the magnetic field and are independent of the MFC in the emitting region. Hence, both of them can not be used as a probe of the MFC in the radiation region.

In addition to the light curves and the energy spectra, polarizations provide two additional observable quantities, i.e., the polarization degree (PD) and polarization angle (PA). The PA rotation could be gradual for the synchrotron radiation in an aligned field, while it would only be abruptly by $90^\circ$ for a toroidal field. Therefore, the PA rotation pattern can be used to test the MFC in the radiation region and the GRB central engine \citep{LWD2016}. Up till now, there has been about 40 bursts with polarization observations in GRB prompt emission \citep{Yonetoku_2011,Yonetoku_2012,Zhang_2019,Sharma_2019,Kole_2020,Chattopadhyay_2022}. Among these polarization observations, there are only 4 GRBs with time-resolved PA analysis during GRB prompt phase and 3 of them show PA rotations with time, i.e., only about $10\%$ of the GRBs show PA rotations in GRB prompt phase. As we know, sufficient photon numbers are needed for the polarization analysis, so the small fraction may be due to the lack of the enough photons for the time-resolved analysis.

GRB 170114A is a single-pulse burst, and the detailed time-resolved polarization analysis shows that its PA rotates abruptly by $\sim$${90}^{\circ}$ twice around the light-curve peak of the prompt phase \citep{Burgess_2019}. The one-emitting-shell model with an aligned field in its radiation region could not interpret such abrupt twice $90^\circ$ PA evolutions \citep{Wang_2023a}. The model with a toroidal field in the emitting region is also problematic because the predicted twice PA rotations are both at the decay phase of the light curves with low flux density \citep{Cheng_2020}. Similar to GRB 170114A, the time-resolved analysis shows that PA of GRB 160821A also changes abruptly by ${90}^{\circ}$ twice with one rotation at the rise phase of the light curve and one at decay phase \citep{Sharma_2019,Chattopadhyay_2022}. The current models could not interpret such observed twice PA rotations.

In this work, we use a multi-emitting-shells model with an aligned field in each shell to explain the twice PA rotations observed in GRB 170114A and GRB 160821A. This paper is arranged as follows. In Section 2, our model is presented. In Section 3, we use our model to interpret multi-windows observations, especially the observed twice abrupt PA rotations of the two bursts. Finally, we make our conclusions and discussion in Section 4.

\section{The Model}\label{models}

Due to the intermittent activity of the central engine, the shells are injected with varying physical properties. The radiation of a burst are composed of these from multiple-emitting-shells. For each shell, it will begin to radiate at a fixed radius $r_{on}$ from the central engine and cease at $r_{off}$. Because of various ejection times, the shells will arrive at $r_{on}$ at a burst source time $t_{on}$, which would be different for each shell. Here, the equal arrival time surface (EATS) effect is considered. For each shell, the time when the photons emitted from the radius $r_{on}$ at a location of $\theta_0$ at burst source time $t_{on}$ are observed is set to be $t_{obs} =t_0$. So photons emitted from a certain location $r$ at a burst-source time $t$ will arrive at the following observational time $t_{obs}$.
\begin{equation}\label{eq:EATS}
t_{obs}-t_0 = [t-\frac{r}{c}\cos\theta-t_{on}+\frac{r_{on}}{c}\cos\theta_0](1+z).
\end{equation}
where $\theta_0$ is zero for on-axis observations, while it equals to $\theta_V-\theta_j$ for off-axis observations. The $\theta_V$ and $\theta_j$ are the observational angle and the jet half-opening angle, respectively. $c$ is the speed of the light and $z$ is the redshift. The $\theta$ is the angle between the local velocity direction and the line of sight in the observer frame.

The radiation model used here for each independent shell is same as that in \cite{Lan_2020,Uhm_2015,Uhm_2016,Uhm_2018}. Each shell will expand radially with relativistic speed. Due to the expansion of the jet shell and the magnetic reconnection process in it, the strength of the magnetic field will decay with radius \citep{Spruit_2001}. 
\begin{equation}\label{eq:mag}
B'(r) = B'_{0} (\frac{r}{r_0})^{-b}.
\end{equation}
The normalization of the magnetic field strength is denoted as $B'_{0}$, $r_0$ is the reference radius and $b$ is the decay index of the magnetic field. The magnetic field in each shell is assumed to be large-scale ordered for a preliminary consideration. The electrons are isotropically injected into the shell and radiate synchrotron emission isotropically in the comoving frame. Therefore, the circular polarization is zero and only the linear polarization is considered \citep{RL1979}.

In the emitting model, the single-energy electrons with Lorentz factor $\gamma_{ch}$ in the comoving frame are assumed. In general, there are two main spectral peak energy ($E_p$) evolution patterns in GRB prompt emission, i.e., the hard-to-soft mode (corresponding to ``i'' model in Uhm et al. 2018) and the intensity-tracking mode (related to ``m'' model in Uhm et al. 2018). For ``i'' model, the variation of the Lorentz factor is a single power law,
\begin{equation}\label{eq:gchi}
\gamma_{ch}(r) = \gamma_{ch}^{0} (\frac{r}{r_0} )^{g},
\end{equation}
The normalization of the electron Lorentz factor for ``i'' model is $\gamma_{ch}^{0}$. For ``m'' model it is a broken power law.
\begin{equation}\label{eq:gchm}
  \gamma_{ch}(r)= \gamma_{ch}^{m}\times\left\{
    \begin{array}{ll}
      (r/r_{m})^{g}, & r\leq r_{m},\\
      (r/r_{m})^{-g}, & r\geq r_{m},
    \end{array}
  \right.
\end{equation}
The normalization of the electron Lorentz factor is $\gamma_{ch}^{m}$ and $r_{m}$ is the corresponding reference radius for ``m'' model. Unless specified, the index $g$ equals to $-0.2$ for ``i'' model and equals to $1$ for ``m'' model.

In the scenario of the magnetic reconnection, the variation of the bulk Lorentz factor of the individual shell is also a single power law with radius.
\begin{equation}\label{eq:gamma}
\Gamma(r) = \Gamma_{0} (\frac{r}{r_0} )^{s},
\end{equation}
The $\Gamma_{0}$ is the normalization value of the bulk Lorentz factor. Depending on the kinds of the central engine, the large-scale ordered MFC in the shell can be aligned (magnetar) or toroidal (black hole). It is predicted that the index $s$ equals to $1/3$ for an aligned configuration, while it is roughly 0 for a toroidal field \citep{Drenkhahn_2002}.

Then the total time-resolved and energy-resolved flux density, the total Stokes parameters Q and U from all the radiation shells of a single burst ($F_{\nu,T}$, $Q_{\nu,T}$ and $U_{\nu,T}$) are expressed as follows.
\begin{eqnarray}
F_{\nu,T}=\sum_{l=1}^{n}  F_{\nu,l} \nonumber\\
Q_{\nu,T}=\sum_{l=1}^{n}  Q_{\nu,l}\\
U_{\nu,T}=\sum_{l=1}^{n}  U_{\nu,l} \nonumber
\end{eqnarray}
n is the total number of the shells. And $F_{\nu,l}$, $Q_{\nu,l}$ and $U_{\nu,l}$ are the flux density, the Stokes parameters Q and U from the ``l''th shell, respectively.  The concrete expressions for $F_{\nu,l}$, $Q_{\nu,l}$ and $U_{\nu,l}$ can be found in the Appendix \ref{app1}.

Then the energy-averaged Stokes parameters are
\begin{eqnarray}\label{stokes}
\bar{f}_\nu=\frac{\int^{\nu_2}_{\nu_1}F_{\nu,T}d\nu}{\int^{\nu_2}_{\nu_1}d\nu} \nonumber\\
\bar{q}_\nu=\frac{\int^{\nu_2}_{\nu_1}Q_{\nu,T}d\nu}{\int^{\nu_2}_{\nu_1}d\nu}\\
\bar{u}_\nu=\frac{\int^{\nu_2}_{\nu_1}U_{\nu,T}d\nu}{\int^{\nu_2}_{\nu_1}d\nu} \nonumber
\end{eqnarray}
The predicted time-resolved and energy-averaged PD ($PD$) and the preliminary PA ($PA_{pre}$) of the emission from one burst with multiple-emitting-shells are expressed as follows.
\begin{equation}
PD=\frac{\sqrt{\bar{q}^2_{\nu}+\bar{u}^2_{\nu}}}{\bar{f}_{\nu}}
\end{equation}
\begin{equation}
PA_{pre}=\frac{1}{2}\arctan\left(\frac{\bar{u}_{\nu}}{\bar{q}_{\nu}}\right)
\end{equation}
For $q_\nu>0$, the final PA ($PA$) equals to $PA_{pre}$. For $q_\nu<0$, the final PA is $PA=PA_{pre}+\pi/2$ when $u_\nu>0$, while it is $PA=PA_{pre}-\pi/2$ when $u_\nu<0$ \citep{Lan2018}.

It should be noted that h$\nu_1=8$ keV and h$\nu_2=40$ MeV for the calculations of the light curves. For the polarization, the values of h$\nu_1$ and h$\nu_2$ in Eq. \ref{stokes} are taken as the concrete values for the corresponding polarization detectors, for example, we take h$\nu_1=50$ keV and h$\nu_2=500$ keV for POLAR's bursts, respectively. The predicted spectral peak energy $E_p$ can be obtained when the value of $\nu\int^{t_2}_{t_1}F_{\nu,T}dt$ reaches its maximum, where $t_1$ and $t_2$ are the begin and end times of the time bins used for obtaining the observed $E_p$ values.

\section{Numerical Results}\label{results}

Observational data sets of the light curves and $E_p$ curves are downloaded in the official web site of Fermi Gamma-ray Burst Monitor (GBM) data~\footnote{https://heasarc.gsfc.nasa.gov/W3Browse/fermi/fermigbrst.html}. A typical Band function is employed in all spectra~\citep{1993BATSE}. All spectral parameters and photon flux of individual spectrum are calculated in the energy band between 8 keV and 40 MeV. The polarization data are obtained from the published papers \citep{Burgess_2019, Sharma_2019}.

For GRB 170114A, both the observed light curve and the spectral peak energy evolution could be roughly fitted by radiation from the single shell. However, it could not interpret the twice PA rotations during the burst \citep{Wang_2023a}. For GRB 160821A, its duration is too long for the single-shell emission. Therefore, we use multi-shells to interpret the light curve, the spectral peak energy evolution, and the polarization curves of the two bursts simultaneously.

Our calculations are based on the formula (7)-(9). The light curve, $E_p$ curve, PD and PA curves are obtained simultaneously. Our fitting processes are as follows: the small bumps in the light curve or the $E_p$ curve would be judged from different shells. The shells are injected from central engine at a discontinuous manner. For each individual shell, it will arrive at $r_{on}$ at a burst source time $t_{on}$. Because of various ejection times of the shells from central engine, $t_{on}$ will be different for each shell, which would correspond to a different observational time $t_0$ of the first arriving photons from each shell. We use a constant time step which equal to 1\% of the $T_{90}$ of the burst, where $T_{90}=T_{95}-T_{5}$ is the burst duration. The accumulated flux reaches 5\% at $T_{5}$ and 95\% at $T_{95}$ of the total flux, respectively. According to the evolution pattern of the spectral peak energy (hard-to-soft or intensity-tracking), the corresponding model (``i'' or ``m'') is used for an individual shell. A schematic figure of our model configuration is shown in Figure 1.

\begin{figure*}[!t]               
\centering
\includegraphics[width=5in]{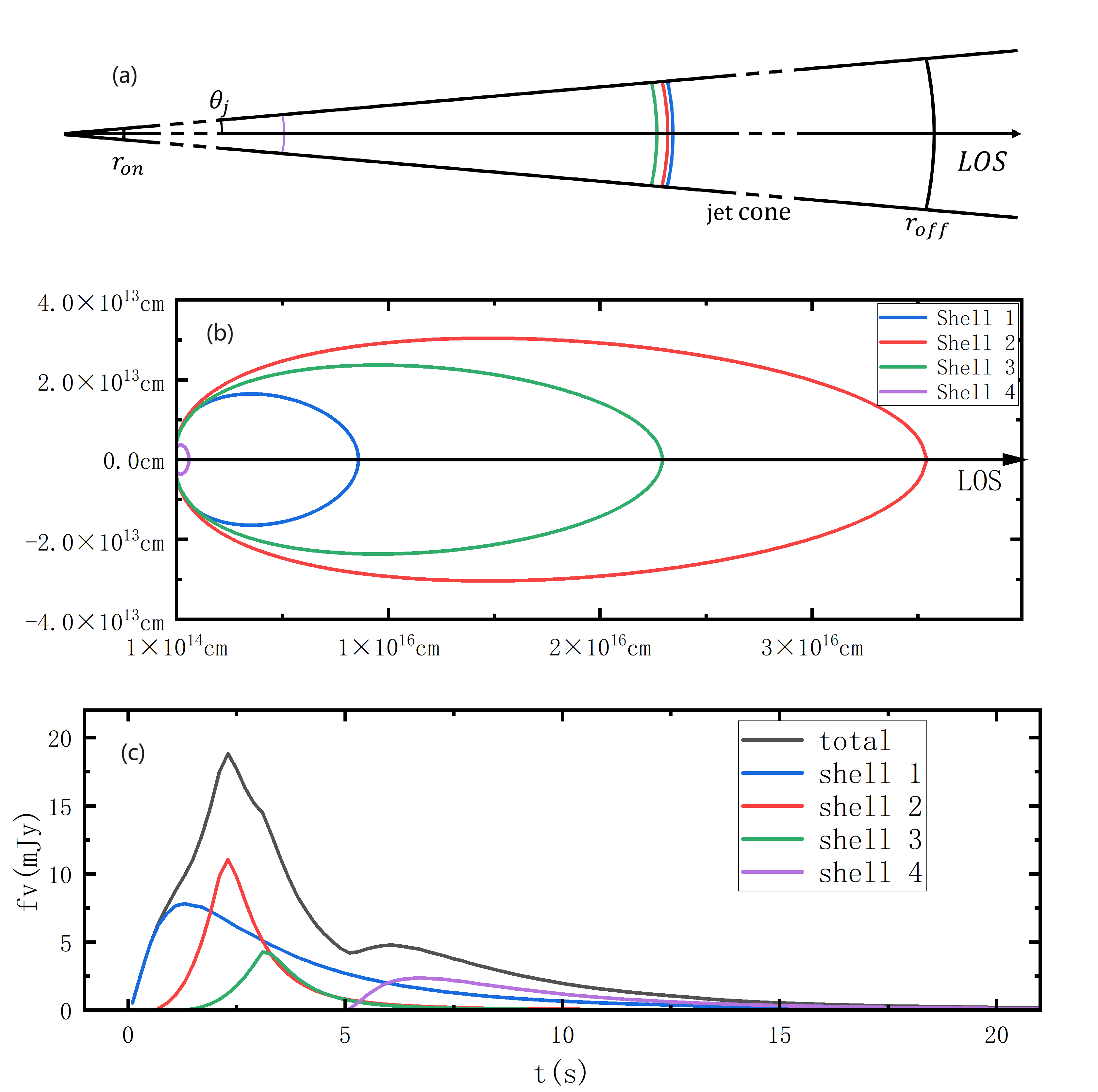}
\caption{\label{fig:diagram}A schematic picture of the model used in Figure \ref{fig:170114A}. The upper panel shows the relative positions of the four shells at a burst source time. The mid panel shows the EATSes for these four shells at a late observational time. Because the bulk Lorentz factors of the four shells within the radiation region ($r_{on}$, $r_{off}$) are different and the EATS would depend on both the observational time and dynamics of the shell, the EATSes of the four shells at the same observational time will be different. The light curves from each shell are shown in the lower panel. The blue, red, green, and purple lines correspond to the shells 1, 2, 3 and 4, respectively.}
\label{fig_sim}
\end{figure*}

There are total 16 (for ``i'' model) or 17 (for ``m'' model) model parameters for each shell. However, with our calculations lots of them show slight effect on our results, we take the typical or fiducial values for these parameters. Finally, there are 7 (i.e., $\Gamma_0$, $r_0$, $R_{inj}$, $t_0$, $\delta$ [for aligned-fields case only], $\theta_V$ and $\gamma_{ch}^0$ for ``i'' model) or 8 (i.e., $\Gamma_0$, $r_0$, $R_{inj}$, $t_0$, $\delta$ [for aligned-fields case only], $\theta_V$, $\gamma_{ch}^m$, and $r_m$ for ``m'' model) model parameters that have significant impacts on our fitting results. For each shell, we first adjust the values of $t_0$, $\Gamma$ ($\Gamma_0$ and $r_0$) and $r_m$ (for ``m'' model only) to match with the observed peak times of the light curve and of the $E_p$ curve. Then the values of $\gamma_{ch}^0$ for the hard-to-soft $E_p$ evolution pattern ($\gamma_{ch}^m$ for the intensity-tracking $E_p$ pattern) is adjusted to match with the observed $E_p$ value and the value of $R_{inj}$ is adjusted to match with the observed flux density. Once the light curve and the $E_p$ curve are both fitted, the key parameters (except $\theta_V$ and $\delta$ [if have]) are almost determined. For the aligned magnetic field, because the predicted PD value for $\theta_V\leq\theta_j$ are similar \citep{SL2024, LWD2021}, we simply take $\theta_V=0$ rad (as shown in Figures 2 and 3). For the toroidal field case, since the PD values are sensitive to $\theta_V$ \citep{SL2024, LWD2021, Toma_2009}, different values of $\theta_V$ are considered in our fitting (see Figures 4 and 5).

Because the decay index of the magnetic field has slight effect on the polarization properties \citep{Lan_2020,Wang_2023a}, we simply set $b=1$. The jet half-opening angle is set as the typical value of $\theta_{j} = 0.1$ rad \citep{Lloyd2019,RE2023}. The values of some parameters we take are the same as these in \cite{Uhm_2018}, i.e., $r_{on} =10^{14}$ cm, and $B'_0 = 30$ G. The normalized bulk Lorentz factor is taken as $\Gamma_{0} = 250$ unless specified. Redshift z is assumed to be 1 for the two bursts because of no reported observation. For the aligned-fields case, the directions of the magnetic fields are set to be rotated by 90° between the shells with $90^\circ$ PA rotation observation. We take $r_{off} = 3\times 10^{16}$ cm for GRB 170114A and $r_{off}=3\times 10^{17}$ cm for GRB 160821A. The values of low- and high-energy spectral index $\alpha$ and $\beta$ are taken as their observational values of the time-integrated ones, i.e., $\alpha=-0.07$ and $\beta=1.04$ for GRB 170114A and $\alpha=-0.01$ and $\beta=1.25$ for GRB 160821A.

For the two bursts, the calculated energy band for the light curve is within (8 keV, 40 MeV), corresponding to the observational energy band of the Fermi/GBM \footnote{\url{https://heasarc.gsfc.nasa.gov/W3Browse/fermi/fermigbrst.html}}. While the time-resolved polarization is calculated within the energy band of POLAR, i.e., (50 keV, 500 keV) for GRB 170114A and within the analysis energy band of AstroSat \citep{Sharma_2019}, i.e., (100 keV, 300 keV) for GRB 160821A.

\begin{figure*}[!t]               
\centering
\includegraphics[width=5in]{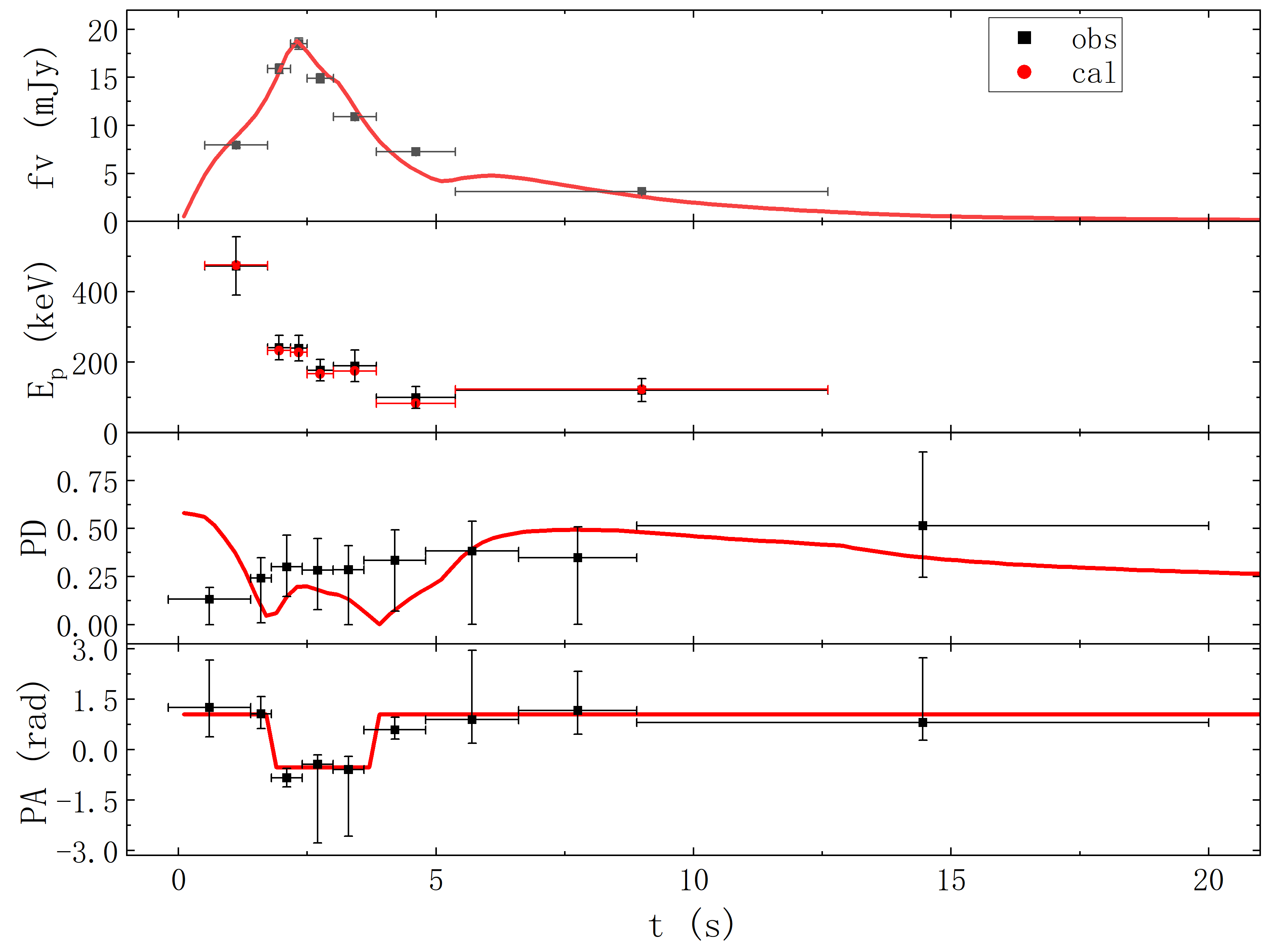}
\caption{\label{fig:170114A}Light curve (top panel), evolution of $E_p$ (second panel), PD curve (third panel), and PA curve (last panel) of GRB 170114A with an aligned magnetic field in each shell. The black squares correspond to observational data. Red circles and lines correspond to our fitting results.}
\label{fig_sim}
\end{figure*}

Since the first PA rotation for GRB 170114A is before the peak time of the light curve of the second shell and the second PA rotation is after the peak time of the light curve of the third shell, the directions of the magnetic fields in these two shells are set to be differently by $90^\circ$ to these of the first and fourth shells to match with the observation. Figure \ref{fig:170114A} shows the fitting result with a large-scale ordered aligned field in each shell and total four shells are used in the fitting. Parameters, which are different between these four shells, are shown in Table 1. Because the magnetic field is assumed to be large-scale ordered, the calculated PD at the first shell is the upper limit, which could be higher than the value of the first observational point. The calculated time-integrated PD is $11.68\%$, which matches the observed one $10.1_{-7.4}^{+10.5} \%$ well.

\begin{figure*}[htbp]               
\centering
\includegraphics[width=5in]{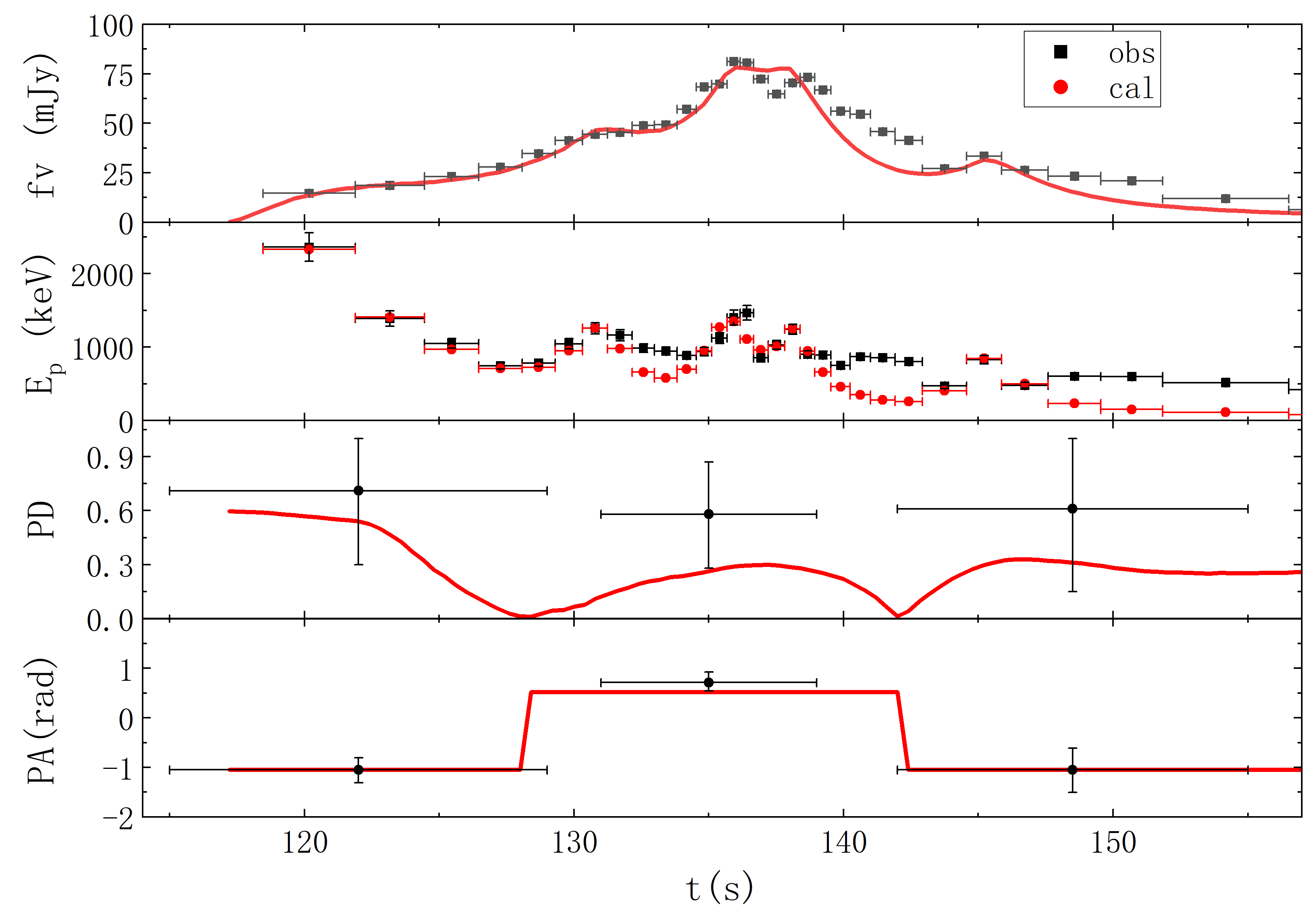}
\caption{\label{fig:160821A}Same as Figure \ref{fig:170114A}, but for GRB 160821A. Polarizations are calculated in the energy band of (100 keV, 300 keV).}
\label{fig_sim}
\end{figure*}

The fitting results with an aligned field in each radiation shell for GRB 160821A are shown in Figure \ref{fig:160821A}. Total five shells are used. Parameters, which are different between these five shells, are shown in Table 2. Within the energy band of 100 to 300 keV, the observed time-integrated PD is $21_{-19}^{+24} \%$ and there are twice 90° PA rotations during the burst. And the calculated time-integrated PD in the same energy band is $4.30\%$ and is consistent with the observed value. The twice PA rotations within the burst duration are reproduced here. Then the analysis energy band of the data was revised to be within (100 keV, 600 keV), an PD upper limit of $33.87\%$ was obtained and the twice 90° PA rotations of the burst was confirmed \citep{Chattopadhyay_2022}. Within (100 keV, 600 keV), the predicted PD is $5.44\%$ and the predicted PA also rotates twice abruptly by $90^\circ$, which are also consistent with the observations.

\begin{figure*}[htbp]
\centering
\includegraphics[width=5in]{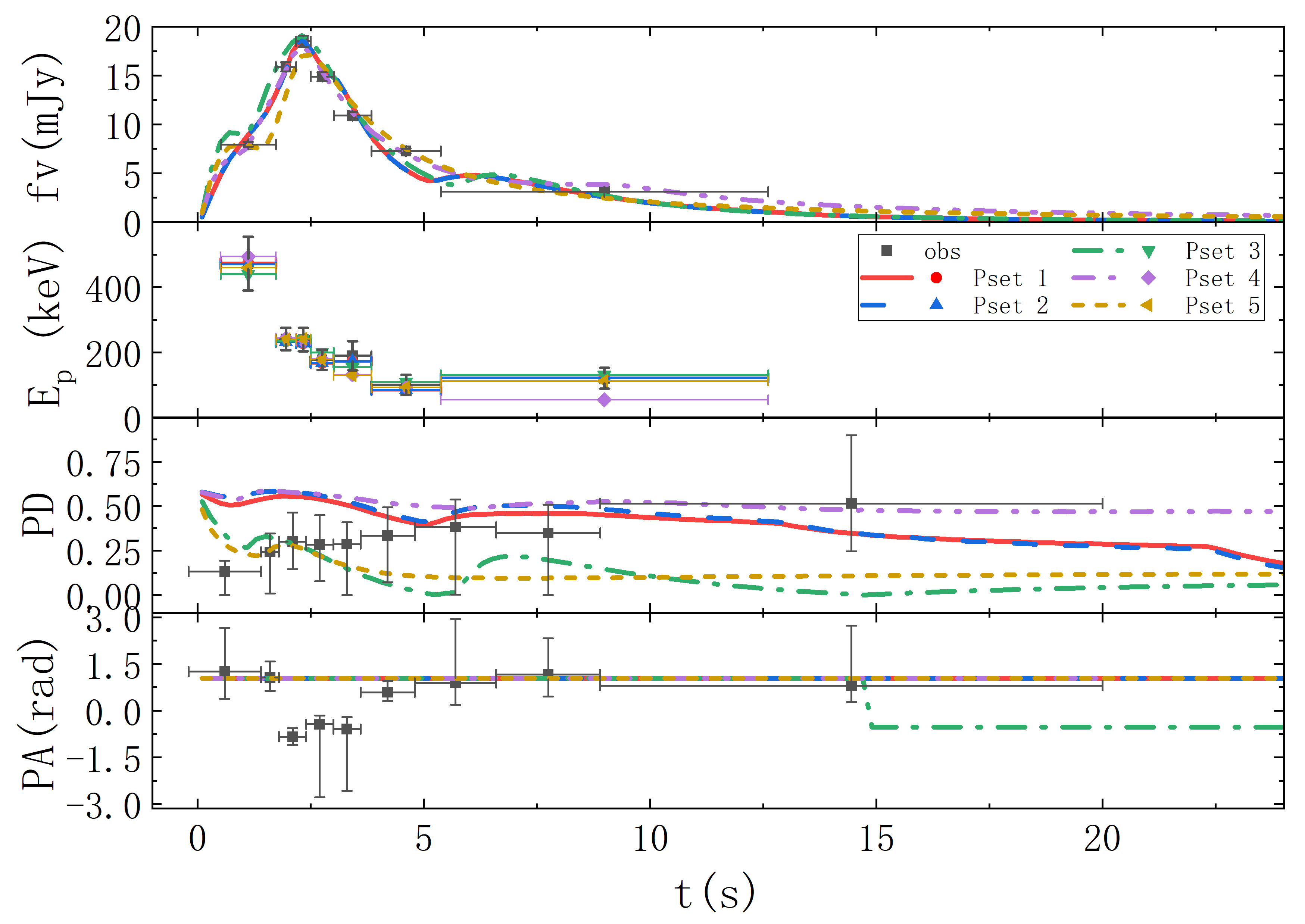}
\caption{\label{fig:170114tors}Light curves (top panel), $E_p$ curves (second panel), PD curves (third panel), and PA curves (last panel) of GRB 170114A with a toroidal magnetic field in each shell. The black squares correspond to the observational data. Five sets of fitting parameters are considered and the corresponding fitting results are shown in color lines and dots. The relevant Parameter sets (Psets) can be found in Table \ref{170114tor}.}
\label{fig_sim}
\end{figure*}

\begin{figure*}[htbp]
\centering
\includegraphics[width=5in]{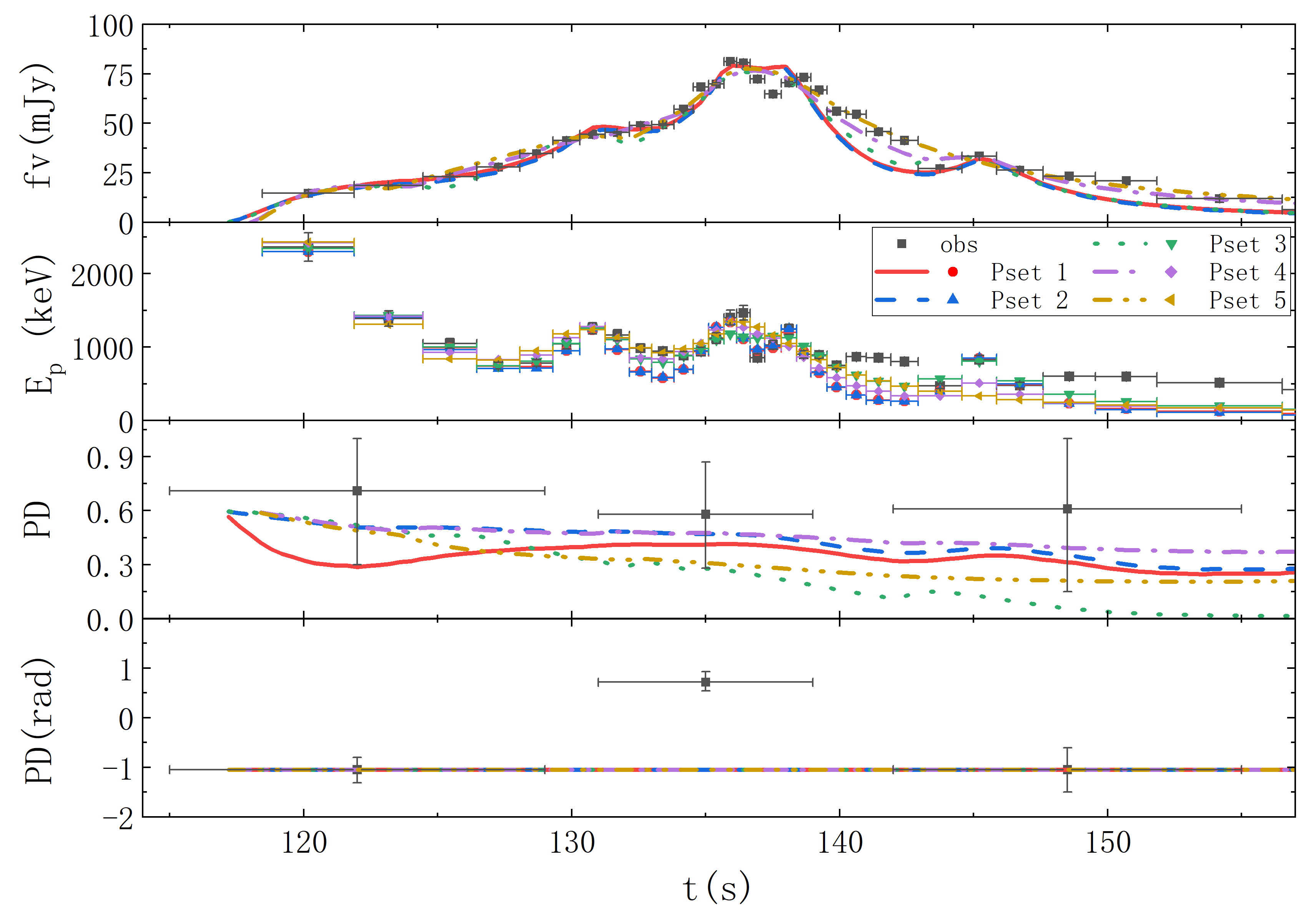}
\caption{\label{fig:160821tors}Same as Figure~\ref{fig:170114tors}, but for GRB 160821A. Polarizations are calculated in the energy band between 100 keV to 300 keV. The relevant Psets can be found in Table \ref{160821tor}.}
\label{fig_sim}
\end{figure*}

For the black hole central engine, the MFCs in the emitting shells are toroidal \citep{Blandford_1977,Spruit_2001}. Although not been predicted, the cases with $s = 0.35$ for the toroidal fields in the emitting regions are also considered to eliminate the influence of the dynamics on the polarizations. Because both the MFC and the value of $q\equiv\theta_V/\theta_j$ with $q\leq1$ have slight effect on the light curves and the evolutions of $E_p$, similar fitting parameters as the aligned-fields case are taken for the cases of $s = 0.35$ with the toroidal fields. The results for the toroidal-fields case are shown in Figures~\ref{fig:170114tors} and~\ref{fig:160821tors}. Although the light curves, $E_p$ curves and PD curves of the two bursts could be fitted by some parameter sets, none of these with toroidal fields in the shells can interpret the observed twice PA rotations during the bursts.

\begin{figure*}[htbp]
\centering
\includegraphics[width=7in]{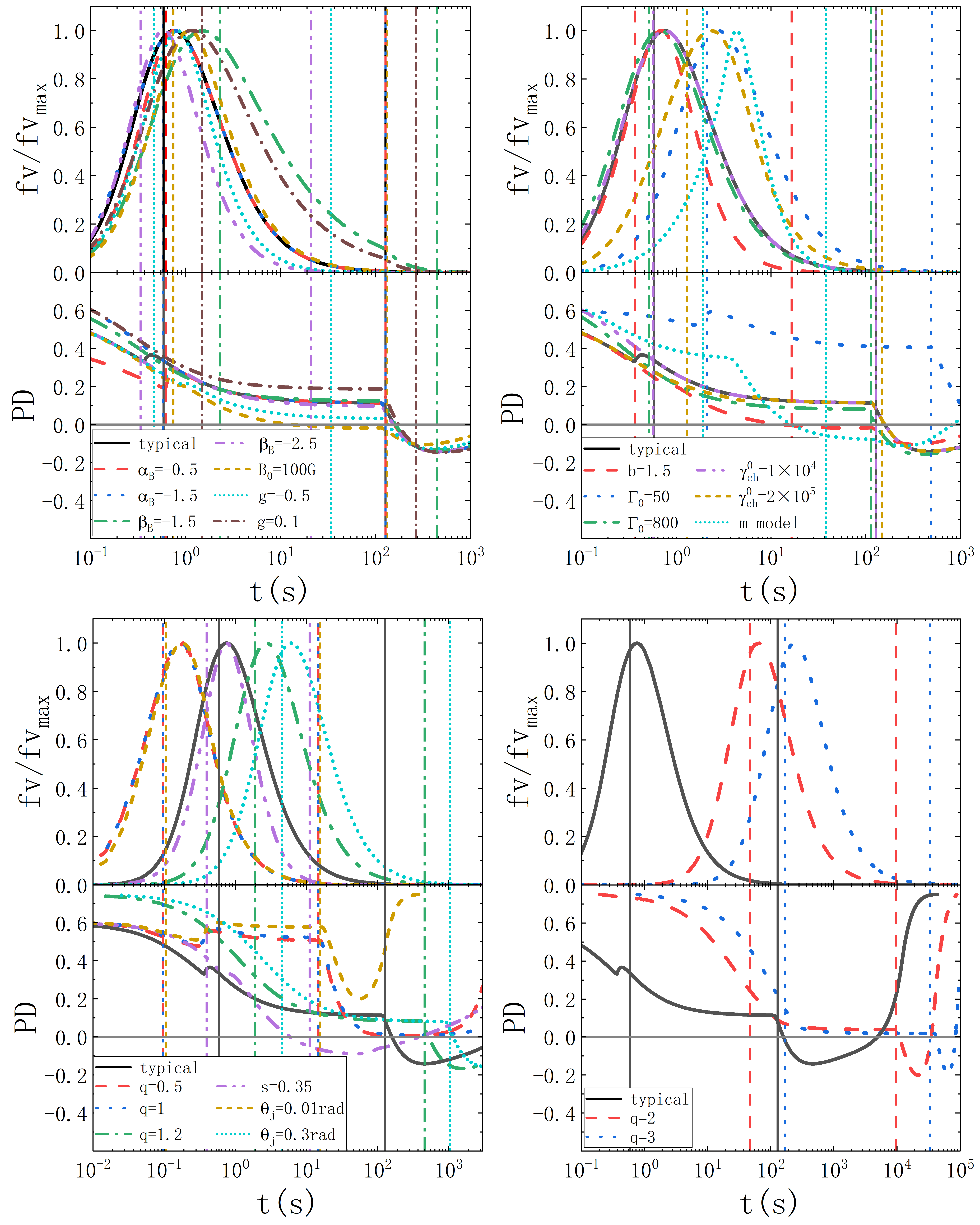}
\caption{\label{fig:typical}Light curves (upper panel) and PD curves (lower panel) for the emission from a single shell with various Psets. A toroidal field is assumed in the emitting shell. The two vertical lines with the same line type as the corresponding light curve show the times of $T_5$ and $T_{95}$ (i.e., the time-accumulated flux reaches 5\% of the total flux at $T_5$ and 95\% of the total flux at $T_{95}$).}
\label{fig_sim}
\end{figure*}

It had been found that PA rotates abruptly during the decay phase of the light curve with a toroidal magnetic field in the radiation region \citep{Cheng_2020}. Here, the PD curves and the light curves from a single emitting shell with a toroidal field, considering various parameter sets, are predicted. The arriving time $t_0$ of the first photons emitted at $r_{on}$ of the shell is set to be 0 for all parameter sets. The typical set of the parameters are set as $s = 0$, $b = 1$, $B'_0 = 30$ G, $\Gamma_0 = 250$, $\theta_j = 0.1$ rad, $q=1.1$, $\alpha = 0.0$, $\beta = 1.0$, $\gamma_{ch}^0 = 5\times10^4$, $g=-0.2$, $r_{0} = 10^{15}$ cm, $r_{on}=10^{14}$ cm, and $r_{off}=3\times10^{16}$ cm. When referring to the ``m'' model, we take $g=1.0$ and $r_{m} = 10^{15}$ cm. The results are shown in Figure~\ref{fig:typical}. In all these cases, PA rotates during the decay phases of the light curves. It's impossible to reproduce the PA rotations around the peak time of the light curve with the toroidal magnetic fields in the emitting regions.

Finally, it should be noted that different value of the bulk Lorentz factor $\Gamma$ for each shell may lead to the collisions of the adjacent shells. Under the parameters we take, the collision radii are all larger than $1.56\times10^{22}$ cm, which is beyond our largest turn-off radius of the radiation ($r_{off}=3\times10^{17}$ cm). Therefore, the collisions would not happen within our calculation radii.

\section{Conclusions and Discussion}\label{sec:Discussion}

In this paper, we use a multi-emitting-shells model to interpret the multi-windows observations, especially the twice $\sim{90}^{\circ}$ PA rotations observed in GRB 170114A and GRB 160821A. The fittings with the aligned magnetic fields in the shells could roughly match the observational data, especially the twice ${90}^{\circ}$ PA rotations of the two bursts, while for the cases with toroidal fields in the shells, the observed PA rotations can not be recovered. However, since we can not calculate all possible cases in the parameter space, the later case can't be ruled out strictly.

Since a toroidal magnetic field in the outflow will correspond to a black hole central engine while an aligned field is usually related to a magnetar \citep{Spruit_2001}, our fitting results show that the central engines for the two bursts are more likely to be magnetars. However, the directions of the magnetic field in the adjacent shells, corresponding to the PA rotation, are required to be differently by $90^\circ$. What mechanism causes the rotation of the magnetic field direction in the adjacent shells is not yet known. One possible process would be a precessing pulsar, i.e., the magnetic axis of the pulsar is inclined an angle with respect to its rotational axis and precesses. And the shells with different orientated magnetic fields are injected out during the precession.

Literally, some works involving the PA evolutions had been done recently \citep{Cheng_2020,Wang_2023a,Wang_2023b}. A large-scale toroidal field was assumed in \cite{Cheng_2020}, and there are indeed two abruptly ${90}^{\circ}$ PA rotations in their model. However, the two PA rotations are both at the decay phase of the light curve. Hence, their model can't interpret the PA rotations at the rise phase. And this result had been confirmed by our calculations with toroidal magnetic fields in the shells. It was predicted that there will be only one ${90}^{\circ}$ PA rotation within $T_{50}$ for the single-shell model with an aligned field for slightly off-axis observations \citep{Wang_2023a}. Therefore, even the field is aligned, a single-shell model also could not interpret the twice ${90}^{\circ}$ PA rotations observed in GRB 170114A and GRB 160821A. Multi-shells should be involved, which indicates an episodic activity of the GRB central engine.

\begin{acknowledgements}

{We thank the anonymous referee for useful comments. We also thank Jin-Jun Geng for useful discussions. We are grateful for the GRB data of Fermi/GBM, POLAR and AstroSat. This work is supported by the National Natural Science Foundation of China (grant Nos. 11903014, 12321003, 12393810, 11833003, 11903017 and 12065017), the National Key Research and Development Program of China (grant no. 2017YFA0402600), the National SKA Program of China No. 2020SKA0120300, International Partnership Program of Chinese Academy of Sciences for Grand Challenges (114332KYSB20210018), Jiangxi Provincial Natural Science Foundation under grant 20224ACB211001 and the Major Science and Technology Project of Qinghai Province (2019-ZJ-A10). M.X.L also would like to appreciate the financial support from Jilin University.}

\end{acknowledgements}

\bibliography{ms_arXiv}
\bibliographystyle{aasjournal}

\appendix

\section{Time-resolved and energy-resolved Stokes parameters from single emitting shell}\label{app1}

Same as in \cite{Lan_2020,Uhm_2015,Uhm_2016,Uhm_2018}, single-energy electrons with Lorentz factor $\gamma_{ch}$ are assumed to mimic the radiation from the true power-law distributed electrons. The spectral power of the radiation from single electron can be divided in to two part, one is the amplitude term $P'_0$ and  the other term, $H_{en}(x)$, describes the shape of the spectrum.
\begin{equation}
P'_{\nu'}(\nu')=P'_0H_{en}(x)
\end{equation}
where $P'_0=3\sqrt{3}m_ec^2\sigma_TB'/(32q_e)$. $m_e$ and $q_e$ are the mass and charge of the electron, $\sigma_T$ represents the Thomson cross section. And $x=\nu'/\nu'_{ch}$ with the observed and critical frequencies in the comoving frame denoted as $\nu'$ and $\nu'_{ch}$, respectively. Then $\nu'=\nu(1+z)/\mathcal{D}$ and $\nu'_{ch}=q_eB'\gamma^2_{ch}\sin \theta'_{B}/2\pi m_ec$. The $\nu$ is the observational frequency and the Doppler factor is denoted as $\mathcal{D}=1/\Gamma/(1-\beta_v\cos\theta)$, where $\beta_v$ is the dimensionless velocity of the shell. The $\theta'_B$ is the pitch angle of the electrons.
The shape term are shown as follows.
\begin{equation}
H_{en}(x)=\begin{cases}
x^{-\alpha}exp(-x), &\text{$x\leq x_c$}, \\ 
x_c^{x_c}exp(-x_c)x^{-\beta}, &\text{$x>x_c$},
\end{cases}
\end{equation}
where $x_c=\beta-\alpha$. And $\alpha$ and $\beta$ are the low- and high-energy spectral index of the Band function \citep{1993BATSE}. 

Then the time-resolved and energy-resolved Stokes parameters of the emission from the single shell can be expressed as followed \citep{Lan_2020}.
\begin{equation}\label{F_v}
F_{\nu,l}=\frac{1+z}{4 \pi D_{L}^{2}} \int \mathcal{D}^{3} \sin \theta d \theta \int  \frac{N P_{0}^{\prime} H_{e n}(x)\sin \theta'_{B}}{4 \pi}d \phi,
\end{equation}
\begin{equation}\label{Q_v}
\begin{split}
Q_{\nu,l}=\frac{1+z}{4 \pi D_{L}^{2}} \int \mathcal{D}^{3} \sin \theta d \theta\times \int \Pi_{p,b} \cos 2 \chi_{p} \frac{N P_{0}^{\prime} H_{e n}(x)\sin \theta'_{B}}{4 \pi}d \phi ,
\end{split}
\end{equation}
\begin{equation}\label{U_v}
\begin{split}
U_{\nu,l}=\frac{1+z}{4 \pi D_{L}^{2}} \int \mathcal{D}^{3} \sin \theta d \theta\times \int  \Pi_{p,b} \sin 2 \chi_{p} \frac{N P_{0}^{\prime} H_{e n}(x)\sin \theta'_{B}}{4 \pi}d \phi.
\end{split}
\end{equation}
where $D_L$ is the luminosity distance and $N=\int{R_{inj} dt}/{\Gamma}$ is the total number of the electrons in the shell. The injection rate of the electrons is $R_{inj}$. The $\phi$ is the angle in the plane of sky between the projection of the jet axis and the projection of the local velocity direction.

The pitch angle $\theta'_{B}$ and the local PA $\chi_{p}$ for the two MFCs can be found in \cite{LWD2016}. And the local PD $\Pi_{p,b}$ used here is a broken power-law \citep{Wang_2023a}.
\begin{equation}
  \Pi_{p,b} = \left\{
      \begin{array}{ll}
      (\alpha+1)/(\alpha+\frac{5}{3}), & x\leq x_c,\\
      (\beta+1)/(\beta+\frac{5}{3}), & x>x_c,
    \end{array}
  \right.
\end{equation}

\section{Parameters used in the fitting}\label{app2}

\begin{table}[h]
 \caption{Fitting parameters of GRB 170114A used in Figure~\ref{fig:170114A}.}
 \label{170114A}
 \begin{tabular*}{\hsize}{@{}@{\extracolsep{\fill}}ccccccc@{}}
  \hline
  \bfseries Shell & \bfseries $t_0 (s)$ & \bfseries $\gamma_{ch}^{0}/\gamma_{ch}^{m}$ & \bfseries $r_0 (cm)$ & \bfseries $r_m (cm)$ & \bfseries $R_{inj} (s^{-1})$ & \bfseries $\delta (rad)$ \\
  \hline
  1 & 0 & $1.45\times 10^4$ & $5.2\times 10^{15}$ & --- & $2.6\times 10^{49}$ & $-\pi/6$\\
  2 & 0.5 & $4.9\times 10^4$ & $2.2\times 10^{15}$ & $1.1\times 10^{15}$ & $8\times 10^{49}$ & $-\pi/6-\pi/2$\\
  3 & 1.2 & $4.85\times 10^4$ & $2.4\times 10^{15}$ & $1.2\times 10^{15}$ & $2.84\times 10^{49}$ & $-\pi/6-\pi/2$\\
  4 & 5 & $6.8\times 10^3$ & $8\times 10^{15}$ & --- & $1.32\times 10^{49}$ & $-\pi/6$\\
  \hline
 \end{tabular*}
\tablecomments{ Other parameters are $\Gamma_{0} = 250$, $s = 0.35$ and $\theta_V = 0$ rad for all these shells.}
\end{table}

\begin{table}[h]
 \caption{Fitting parameters of GRB 160821A used in Figure~\ref{fig:160821A}.}
 \label{160821A}
 \begin{tabular*}{\hsize}{@{}@{\extracolsep{\fill}}ccccccc@{}}
  \hline
  \bfseries Shell & \bfseries $t_0 (s)$ & \bfseries $\gamma_{ch}^{0}/\gamma_{ch}^{m}$ & \bfseries $r_0 (cm)$ & \bfseries $r_m (cm)$ & \bfseries $R_{inj} (s^{-1})$ & \bfseries $\delta (rad)$ \\
  \hline
  1 & 117 & $8.8\times 10^3$ & $5.5\times 10^{16}$ & --- & $2\times 10^{48}$ & $\pi/6$\\
  2 & 122 & $8.1\times 10^4$ & $1.5\times 10^{16}$ & $1.9\times 10^{15}$ & $7.2\times 10^{48}$ & $\pi/6+\pi/2$\\
  3 & 130 & $7.1\times 10^4$ & $1.35\times 10^{16}$ & $1\times 10^{15}$ & $1.32\times 10^{49}$ & $\pi/6+\pi/2$\\
  4 & 134 & $9.5\times 10^4$ & $7.8\times 10^{15}$ & $1\times 10^{15}$ & $1.2\times 10^{49}$ & $\pi/6+\pi/2$\\
  5 & 140 & $6.5\times 10^4$ & $1.1\times 10^{16}$ & $1.1\times 10^{15}$ & $8.4\times 10^{48}$ & $\pi/6$\\
  \hline
 \end{tabular*}
\tablecomments{ Other parameters are $\Gamma_{0} = 250$, $s = 0.35$ and $\theta_V = 0$ rad for all these shells.}
\end{table}

\begin{table}[h]

\caption{Parameter sets of GRB 170114A used in Figure~\ref{fig:170114tors}.}

\label{170114tor}

\begin{tabular*}{\hsize}{@{}@{\extracolsep{\fill}}c|c|ccccccc@{}}

\hline

\bfseries Pset & \bfseries $q$ & \bfseries Shell & \bfseries $t_0 (s)$ & \bfseries $\Gamma_{0}$ & \bfseries $\gamma_{ch}^{0}/\gamma_{ch}^{m}$ & \bfseries $r_0 (cm)$ & \bfseries $r_m (cm)$ & \bfseries $R_{inj} (s^{-1})$ \\

\hline
1&0.2&1 & 0 & 250 & $1.45\times 10^4$ & $5.2\times 10^{15}$ & --- & $2.6\times 10^{49}$\\
&&2 & 0.5 & 250 & $4.9\times 10^4$ & $2.2\times 10^{15}$ & $1.1\times 10^{15}$ & $8\times 10^{49}$\\
&&3 & 1.2 & 250 & $4.85\times 10^4$ & $2.4\times 10^{15}$ & $1.2\times 10^{15}$ & $2.84\times 10^{49}$\\
&&4 & 5 & 250 & $6.8\times 10^3$ & $8\times 10^{15}$ & --- & $1.32\times 10^{49}$\\
\hline
2&0.6&1 & 0 & 250 & $1.45\times 10^4$ & $5.2\times 10^{15}$ & --- & $2.6\times 10^{49}$\\
&&2 & 0.5 & 250 & $4.9\times 10^4$ & $2.2\times 10^{15}$ & $1.1\times 10^{15}$ & $8\times 10^{49}$\\
&&3 & 1.2 & 250 & $4.85\times 10^4$ & $2.4\times 10^{15}$ & $1.2\times 10^{15}$ & $2.84\times 10^{49}$\\
&&4 & 5 & 250 & $6.8\times 10^3$ & $8\times 10^{15}$ & --- & $1.32\times 10^{49}$\\
\hline
3&1.1&1 & 0 & 250 & $6.2\times 10^4$ & $1\times 10^{15}$ & --- & $7.2\times 10^{51}$ \\
&&2 & 1 & 250 & $8.5\times 10^4$ & $1\times 10^{15}$ & $3.3\times 10^{14}$ & $2.4\times 10^{52}$ \\
&&3 & 5.5 & 250 & $1.3\times 10^4$ & $5\times 10^{15}$ & --- & $2.4\times 10^{50}$ \\
\hline
4&0.6&1 & 0 & 70 & $4.6\times 10^4$ & $1\times 10^{15}$ & --- & $1.8\times 10^{50}$ \\
&&2 & 1 & 100 & $6.3\times 10^4$ & $1\times 10^{15}$ & $4.5\times 10^{14}$ & $3.48\times 10^{50}$ \\
&&3 & 6 & 100 & $8.5\times 10^4$ & $1\times 10^{15}$ & $1\times 10^{15}$ & $4\times 10^{49}$ \\
\hline
5&1.1&1 & 0 & 200 & $7.5\times 10^4$ & $1\times 10^{15}$ & --- & $1.72\times 10^{52}$ \\
&&2 & 1.4 & 250 & $8.8\times 10^4$ & $1\times 10^{15}$ & $2.8\times 10^{14}$ & $1.08\times 10^{53}$ \\
\hline

\end{tabular*}

\tablecomments{ s is 0 for the last two sets of parameters, while it is 0.35 for other parameter sets.}

\end{table}

\begin{table}[h]
 \caption{Parameter sets of GRB 160821A used in Figure~\ref{fig:160821tors}. }
 \label{160821tor}
 \begin{tabular*}{\hsize}{@{}@{\extracolsep{\fill}}c|c|ccccccc@{}}
  \hline
  \bfseries Pset & \bfseries $q$ & \bfseries Shell & \bfseries $t_0 (s)$ & \bfseries $\Gamma_{0}$ & \bfseries $\gamma_{ch}^{0}/\gamma_{ch}^{m}$ & \bfseries $r_0 (cm)$ & \bfseries $r_m (cm)$ & \bfseries $R_{inj} (s^{-1})$ \\
  \hline
  1&0.2&1 & 117 & 250 & $8.8\times 10^3$ & $5.5\times 10^{16}$ & --- & $2\times 10^{48}$\\
  &&2 & 122 & 250 & $8.1\times 10^4$ & $1.5\times 10^{16}$ & $1.9\times 10^{15}$ & $7.2\times 10^{48}$\\
  &&3 & 130 & 250 & $7.1\times 10^4$ & $1.35\times 10^{16}$ & $1\times 10^{15}$ & $1.32\times 10^{49}$\\
  &&4 & 134 & 250 & $9.5\times 10^4$ & $7.8\times 10^{15}$ & $1\times 10^{15}$ & $1.2\times 10^{49}$\\
  &&5 & 140 & 250 & $6.5\times 10^4$ & $1.1\times 10^{16}$ & $1.1\times 10^{15}$ & $8.4\times 10^{48}$\\
  \hline
  2&0.6&1 & 117 & 250 & $8.8\times 10^3$ & $5.5\times 10^{16}$ & --- & $2\times 10^{48}$\\
  &&2 & 122 & 250 & $8.1\times 10^4$ & $1.5\times 10^{16}$ & $1.9\times 10^{15}$ & $7.2\times 10^{48}$\\
  &&3 & 130 & 250 & $7.1\times 10^4$ & $1.35\times 10^{16}$ & $1\times 10^{15}$ & $1.32\times 10^{49}$\\
  &&4 & 134 & 250 & $9.5\times 10^4$ & $7.8\times 10^{15}$ & $1\times 10^{15}$ & $1.2\times 10^{49}$\\
  &&5 & 140 & 250 & $6.5\times 10^4$ & $1.1\times 10^{16}$ & $1.1\times 10^{15}$ & $8.4\times 10^{48}$\\
  \hline
   3&1.1&1 & 117 & 250 & $1\times 10^4$ & $4.8\times 10^{16}$ & --- & $1.16\times 10^{49}$ \\
   &&2 & 125 & 250 & $9.5\times 10^4$ & $1\times 10^{16}$ & $6.5\times 10^{14}$ & $1.96\times 10^{50}$ \\
   &&3 & 132 & 250 & $1.6\times 10^5$ & $4\times 10^{15}$ & $6\times 10^{14}$ & $8.8\times 10^{50}$ \\
   &&4 & 134 & 250 & $1.03\times 10^5$ & $6.8\times 10^{15}$ & $5\times 10^{14}$ & $3.72\times 10^{50}$ \\
   &&5 & 142 & 250 & $1.35\times 10^5$ & $3.5\times 10^{15}$ & $5\times 10^{14}$ & $6\times 10^{50}$ \\
  \hline
   4&0.6&1 & 118 & 33 & $1.4\times 10^5$ & $1\times 10^{15}$ & --- & $1.6\times 10^{50}$ \\
   &&2 & 124 & 100 & $3.5\times 10^5$ & $1\times 10^{15}$ & $2\times 10^{15}$ & $1.48\times 10^{50}$ \\
   &&3 & 131 & 100 & $3.4\times 10^5$ & $1\times 10^{15}$ & $1.5\times 10^{15}$ & $1.52\times 10^{50}$ \\
   &&4 & 134 & 100 & $2.5\times 10^5$ & $1\times 10^{15}$ & $1.2\times 10^{15}$ & $1.6\times 10^{50}$ \\
   &&5 & 143 & 100 & $2.5\times 10^5$ & $1\times 10^{15}$ & $7.4\times 10^{14}$ & $4\times 10^{49}$ \\
  \hline
   5&1.1&1 & 118 & 40 & $1.45\times 10^5$ & $1\times 10^{15}$ & --- & $6.8\times 10^{50}$ \\
   &&2 & 123 & 100 & $3.8\times 10^5$ & $1\times 10^{15}$ & $1.1\times 10^{15}$ & $3.08\times 10^{51}$ \\
   &&3 & 132 & 100 & $3.3\times 10^5$ & $1\times 10^{15}$ & $6.2\times 10^{14}$ & $3.08\times 10^{51}$ \\
   &&4 & 134 & 100 & $2.8\times 10^5$ & $1\times 10^{15}$ & $6\times 10^{14}$ & $1.6\times 10^{51}$ \\
  \hline
 \end{tabular*}
\tablecomments{ s is 0 for the last two sets of the parameters, while it is 0.35 for other parameter sets.}
\end{table}

\end{document}